\documentclass[aps, pre, twocolumn, a4paper, groupedaddress, floatfix]{revtex4}
\usepackage{graphicx}
\usepackage{amsmath}
\usepackage{placeins}
\usepackage{color}
\usepackage{soul}
\usepackage{hyperref}
\usepackage{textcomp}

\begin{document}

\title{Two Price Regimes in Limit Order Books:\\Liquidity Cushion and Fragmented Distant Field}

\author{Sebastian M. Krause, Edgar Jungblut, Thomas Guhr}
\affiliation{Fakultät für Physik, Universität Duisburg-Essen, Duisburg, Germany}

\begin{abstract}

The distribution of liquidity within the limit order book is essential for the impact of market orders on the stock price and the emergence of price shocks. Limit orders are characterized by stylized facts: The number of inserted limit orders declines with the price distance from the quotes following a power law and limit order lifetimes and volumes are power law distributed. Strong dependencies among these quantities add to the complexity of limit order books. Here we analyze the limit order book in the dimensions of price, time, limit order lifetime and volume altogether. This allows us to identify regularities which are not visible in marginal distributions. Particularly we find that the limit order book is divided into two regimes. Around the quotes we find a densely filled regime with mostly short living limit orders closely adapting to the price. Far away from the quotes we find a sparse filling with long living limit orders, mostly inserted at particular times of the day being prone to flash crashes. We determine the characteristics of those two regimes and point out the main differences. Based on our research we propose a model for simulating the regime around the quotes.

\end{abstract}
 
\maketitle

\section{Introduction}

Setting up consistent models which are capable of capturing and describing the whole variety of empirical findings in large complex systems is a formidable challenge, requiring a lot of staying power and willingness to constantly reiterate. A major reason for this is non-stationarity, \textit{i.e.} changes of crucial system features which are often seemingly erratic. Thus, large-scale data analysis is called for to identify universal or generic features as corner stones for model building.

Here, we want to contribute to this enterprise in the case of financial markets. A wealth of data is available. Schematic models such as stochastic processes are highly important, but have their clear limitation in view of the microscopic dynamics governed by the order book. Agent based models often suffer from too many parameters of unclear meaning or, more generally, merely mimic the dynamics of the trading without deeper insight to the mechanisms. This applies in particular to the closing and opening of markets. Thus, careful data analyses are needed to provide better quantitative information on the order book dynamics which can then be used to pave the road to improved agent based models with a considerably lower number of parameters.

Financial markets are highly non-stationary systems \cite{Mandelbrot.1997b, Mantegna.1997, Munnix.2012}. Market conditions are constantly adapting to technological progress and regulatory changes \cite{Chung.2016, Breuer.2020, Hendershott.2013}. Detailed analysis of order flow data describing order book dynamics is called for to unravel the various governing influences from the viewpoint of complex systems. In addition such an analysis may shed light on systemic issues such as the benefits of (de)regulations \cite{Farmer.2009} and the impact of high frequency trading on market stability \cite{Lee.2013, KIRILENKO.2017, Braun.2018}. In this context an improvement of agent based models based on the analysis of order flow dynamics is desirable \cite{Patzelt.2013, Meudt.2016, Schmitt.2012, Krause.2013}. Stylized facts are well-established for price time series \cite{Plerou.1999, Cont.2001} and highly robust over time as well as for different market places. Therefore they can coherently be included into risk modeling. On the other hand, stylized facts known for order book dynamics differ between market places and change their manifestation over time \cite{Gould.2013, Theissen.2017}. For example, the price distance of inserted limit orders from the quotes is known to follow a power law \cite{Bouchaud.2002, Potters.2003}. The broad distribution can be interpreted as a consequence of market participants expecting large price changes at any moment \cite{Bouchaud.2002, Potters.2003}. Further, the lifetime of limit orders until being traded or canceled plays a key role in the build-up of liquidity \cite{Challet.2001}. It can be understood as an interplay of a first passage process in the presence of diffusive price movement, together with a power law for cancellation times \cite{Eisler.2009}. The distribution of liquidity within the limit order book is crucial for the stability of the limit order book and the response to large trades \cite{Schmitt.2012}. In view of the non-stationarity, the data analysis has to be redone often. For example, NASDAQ stocks showed a lot of orders being placed far from the quotes in 2002 \cite{Potters.2003}, but later studies found many orders being placed near the quotes, even suggesting queue models for the characterization of their execution sequence \cite{Cont.2010, Gareche.2013, Huang.2015}.

Another important reason why stylized facts of limit order books are more puzzling than those of price dynamics is the following: Quantities as limit order lifetime and insertion price strongly depend on each other \cite{Eisler.2009}. This has to be taken into account for agent based models, for example with a multivariate distribution of order lifetimes and insertion prices \cite{Eisler.2009}. The marginal distributions are not enough to fully comprehend the filling process of the limit order book.

Here we go one step further and analyze the limit order book in the dimensions of price, time, order lifetime and order volume altogether. This allows us to identify regularities which are not visible in the marginal distributions. We analyze order flow data of 96 NASDAQ stocks in early 2016. For all stocks we find that the order book is divided into two regimes. These regimes seem to be connected to two tendencies which come along with automated trading: On the one hand the spread and the associated cost of trading decrease for frequently traded stocks. On the other hand the limit order book is filled up deep, which might be a strategy to utilize or depress flash crashes. This leads to a rich, staggered limit order book structure with different time scales for the dynamics. Inspired by the visual appearance of the time dependent limit order book, we find a quantity allowing us to identify the size of the densely filled regime close to the quotes. This procedure works for all 96 stocks in an automated fashion. The marginal distributions being mostly analyzed in the literature are unsuitable for this purpose. We find that the regimes of the limit order book are ruled by very different dynamic processes calling for different types of agent based models. For the regime around the quotes we analyze an agent based model and find that more details than in \cite{Bouchaud.2002} have to be included, while a detailed model as in \cite{Eisler.2009} is not necessary to understand the filling process of the limit order book.

Our study is organized as follows: In Sec.~\ref{Data} we describe the data. In Sec.~\ref{Data analysis} we reconstruct the limit order book and investigate its progression during the trading day. In Sec.~\ref{Model} we propose a model for simulating the regime around the quotes and compare it to the market data. In Sec.~\ref{Conclusion} we summarize our results and provide an outlook.

\section{Data}
\label{Data}

We analyze stocks from NASDAQ US with the Historical TotalView-ITCH data, downloaded from \cite{tradingphysics}. The data ranges over the working week from March 7, 2016 to March 11, 2016 and comprises 96 stocks of the 100 stocks listed in the NASDAQ 100 at this time, as four stocks are not available. The data contains information about limit order book changes during a trading day, \textit{i.e.} limit order insertions, (partial) cancellations and (partial) trades. Moreover, they contain information on trades against hidden orders. The specifications for \textit{e.g.} each limit order insertion $i$ include among other things the insertion time $t_i$, the volume/number of shares $v_i$ and the price $p_i$ of the order. A detailed description of the data can be found in \cite{Huang.2011, tradingphysics_specs}.

We analyze data for a whole trading day from 4:00 am to 8:00 pm (New York time) with a focus on the regular trading times of the NASDAQ, 9:30 am to 4:00 pm (New York time) \cite{nasdaq_hours}. All events have a time stamp in milliseconds. Multiple events may have the same time stamp. For those events chronological order is maintained because incoming orders are processed by the market as they arrive.

We reconstruct the order book volume $v(p,t)$ of all visible orders for each price level $p$ and each time $t$ of the trading day for every stock on every trading day. In this way we can infer the best available ask price bestask$(t)$ and the best available bid price bestbid$(t)$ for every time $t$. The difference between those two is the spread
\begin{align}
s(t) =& \textrm{bestask}(t) – \textrm{bestbid}(t).
\end{align}
It is an important measure for the market liquidity, a small spread implies a low cost for trading. Moreover, the overall structure of the orderbook is found to be highly dependent on the size of the spread \cite{Theissen.2017}. Stocks with a spread as small as one tick are characterized as large-tick stocks with a densely filled order book around the quotes and a sparsely filled order book far from the quotes. A large spread often implies gaps behind the quotes and a more filled order book far from the quotes. We characterize the price of the stock by the midpoint price
\begin{align}
m(t) =&\frac{ \textrm{bestask}(t) + \textrm{bestbid}(t)}{2}.
\end{align}
This is a measure independent of buying or selling the stock.

\section{Data analysis}
\label{Data analysis}

In Sec.~\ref{Time evolution}, we briefly describe the time evolution of the limit order book. In Sec.~\ref{Lifetimes}, we analyze in depth the distribution of the limit order lifetimes and volumes. In Sec.~\ref{Cushion}, we determine the width of the densely filled regime around the quotes. In Sec.~\ref{CushionComp}, we examine the relationship between the width of the densely filled regime and other trading characteristics. In Sec.~\ref{distant field}, we point out the main differences between limit order insertions close to the quotes and farther away from the quotes.

\subsection{Time evolution of the limit order book}
\label{Time evolution}

\begin{figure}[htb]
\begin{center}
	\includegraphics[width=1.0\columnwidth]{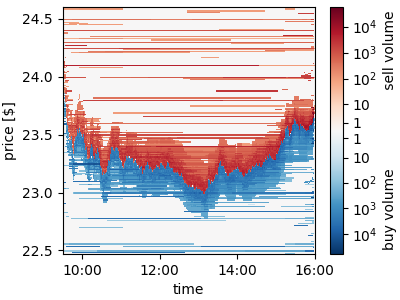}
	\caption{Color plot of the time evolution of volume in the limit order book over a section of limit order prices around the midpoint price. Volumes are shown on a logarithmic color scale (red for sell limit orders, blue for buy limit orders and light gray for empty levels) for \textsc{Ebay} on March 10, 2016.}
	\label{fig:LOB_Ebay}
\end{center}
\end{figure}

\begin{figure}[htb]
	\begin{center}
		\includegraphics[width=1.0\columnwidth]{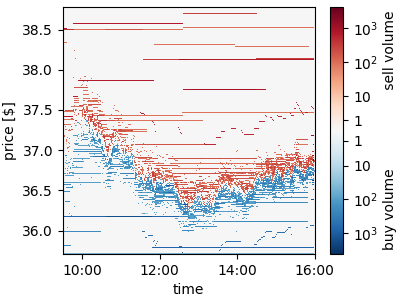}
		\caption{Color plot of the time evolution of volume in the limit order book over a section of limit order prices around the midpoint price. Volumes are shown on a logarithmic color scale (red for sell limit orders and blue for buy limit orders) for \textsc{Liberty Interactive Corporation} on March 10, 2016. }
		\label{fig:LOB_LVNTA}
	\end{center}
\end{figure}

\begin{figure*}[htb]
	\begin{center}
		\includegraphics[width=0.98\textwidth]{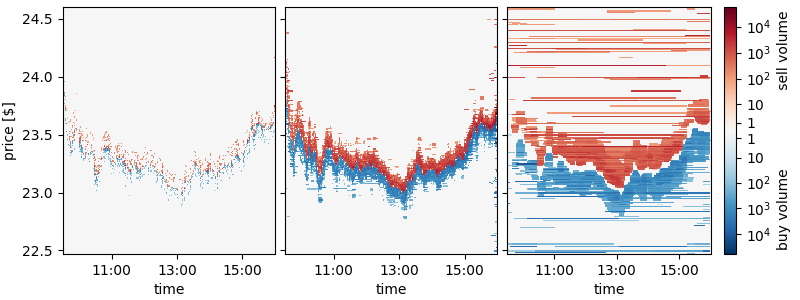}
		\caption{Left: Limit order book as in Fig.~\ref{fig:LOB_Ebay} reduced only to short-living limit orders with a lifetime $\tau_i < 10\,$s. Middle: Limit order book of limit orders with a lifetime $10\,$s $\leq \tau_i < 10\,$min. Right: Limit order book of long-living limit orders with lifetimes $\tau_i \geq 10\,$min.}
		\label{fig:LOB_Ebay_lifetimes}
	\end{center}
\end{figure*}

The time evolution of the limit order book for the stock of \textsc{Ebay} in Fig.~\ref{fig:LOB_Ebay} shows the occupation of the limit order book price niveaus versus time for a whole trading day \cite{Todd.2014, Bookstaber.2015, Sandoval.2016}. The midpoint price is seen to be surrounded by a densely filled regime with a width of about 50 cents which is continuously adapted to midpoint price changes, and a more static fragmented regime far from the midpoint price. There is much more volume at prices farther from the midpoint price, for buy limit orders at lower prices down to one cent and for sell limit orders at larger prices up to 199999\,\$. Between best sell and best buy price there is no gap during most of the time, the stock has the smallest possible spread of one cent and therefore we are dealing with a so called large-tick stock. 

The behavior of a small-tick stock, a stock with a spread large in comparison to the tick size, is similar, as shown in Fig.~\ref{fig:LOB_LVNTA}. Close to the midpoint price there is a fast adapting regime which is more densely filled than the limit order book far from the midpoint price. The large spread implies a gap between best sell and best buy offer, and behind these levels there are gaps. The dynamics of the order book of small-tick stocks have been analyzed in \cite{Schmitt.2012}.

\begin{figure}[htb]
	\begin{center}
		\includegraphics[width=0.98\columnwidth]{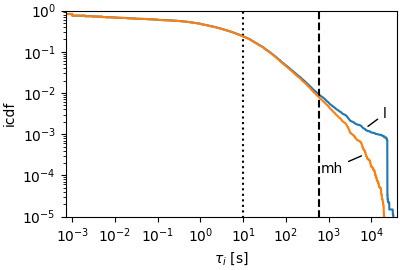}
		\caption{Inverse cumulative distribution function of limit order lifetimes during the whole day (blue line, `d') and during market hours between 9:30 and 16:00 (orange line, `mh') on a double-logarithmic scale for \textsc{Ebay} on March 10, 2016. The dotted vertical line highlights ten seconds and the dashed vertical line ten minutes.}
		\label{fig:lifetimes}
	\end{center}
\end{figure}

Studies on NASDAQ \cite{Potters.2003}, the Paris Bourse \cite{Bouchaud.2002} and the Shenzhen Stock Exchange \cite{Gu.2008b} showed a similar dense filling of the order book close to the quotes \cite{Gould.2013}.

\subsection{Limit order lifetimes}
\label{Lifetimes}

\begin{figure}[htb]
	\begin{center}
		\includegraphics[width=0.98\columnwidth]{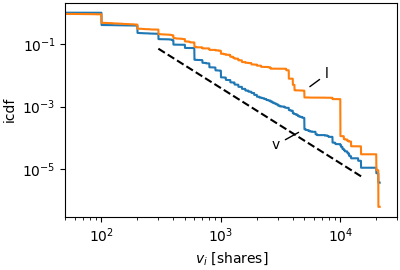}
		\caption{Inverse cumulative distribution function of limit order volumes (blue line, `v') and limit order volumes weighted with their lifetimes (orange line, `l') on a double-logarithmic scale for \textsc{Ebay} on March 10, 2016. The dashed black line is a power law $v_i^{-2.4}$.}
		\label{fig:weighted_volume}
	\end{center}
\end{figure}

\begin{figure}[htb]
	\begin{center}
		\includegraphics[width=0.98\columnwidth]{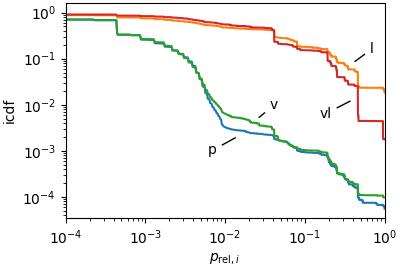}
		\caption{Inverse cumulative distribution function of relative insertion prices (blue line, `p'), weighted with the volume (green line, `v'), weighted with the lifetime (orange line, `l') and weighted with the product of volume and lifetime (red line, `vl') on a double-logarithmic scale for \textsc{Ebay} on March 10, 2016.}
		\label{fig:weighted_price}
	\end{center}
\end{figure}

The availability of large volumes in the limit order book is important for the liquidity of a market. It is therefore worth studying how the volume is provided - with a large number of short-lived limit orders or with a small number of long-lived orders? We determine the lifetime $\tau_i$ of a limit order as the time since it is inserted until it is traded or cancelled.
To understand how limit orders of short lifetimes contribute to filling the limit order book, we erase all limit orders with a lifetime $\tau_i \geq 10\,$s and reconstruct a reduced limit order book of short-living orders. The result is shown in Fig.~\ref{fig:LOB_Ebay_lifetimes} on the left. Although the short-living orders are a large majority with a share of 76\,\% among all limit orders, they only contribute to a sparse filling of the limit order book. As seen in the middle of Fig.~\ref{fig:LOB_Ebay_lifetimes}, the contribution of orders with medium lifetimes $10\,$s $\leq \tau_i < 10\,$min is much larger and provides almost the complete volume around the quotes. Long-living orders with lifetimes $\tau_i \geq 10\,$min provide volumes farther from the quotes, as Fig.~\ref{fig:LOB_Ebay_lifetimes} on the right illustrates. Although the share of these orders is less than 1\,\%, they are responsible for almost all the volume in the depth of the limit order book. 

Order lifetimes range from milliseconds to hours.
In Fig.~\ref{fig:lifetimes} the inverse cumulative distribution function (icdf)
\begin{equation}
I(x) = 1 - F(x) = 1 - \int\limits_{-\infty}^{x} f(x') dx' ,
\label{eq:icdf}
\end{equation}
also referred to as quantile function, of order lifetimes reveals that they are broadly distributed. This is in accordance with previous studies, where the lifetime distribution was found to be fat tailed and described by a power law \cite{Eisler.2009, Challet.2001}. Out of all limit orders 0.7\,\textperthousand \ live the whole trading day (6.5\,h or $2.34\cdot 10^4$\,s). A small fraction of these orders lives even longer, as they are introduced already in the pre-market hours or continue in the after-market. Out of all limit orders 0.9\,\% live at least ten minutes (600\,s) and 24\,\% live at least ten seconds. Limit orders which are inserted and cancelled or traded within the same millisecond have an unknown lifetime of less than one millisecond due to the limited time resolution. The share of these limit orders is 17\,\%, therefore the icdf reaches a value of 83\,\% at one millisecond. Only including limit orders during the trading hours between 9:30 and 16:00, the according icdf is similar to the icdf for limit orders over the whole day. Differences appear only for long lifetimes $\tau_i \geq 10\,$min. 

Combining the fact that the distribution of order lifetimes spans many orders of magnitude (Fig.~\ref{fig:lifetimes}) with the finding that orders of various lifetimes have very different contributions to the limit order book filling (Fig.~\ref{fig:LOB_Ebay_lifetimes}), we conclude that statistical properties of limit orders have to be analyzed with awareness of order lifetimes. In the literature, probability densities of limit order properties as volume or insertion price are often calculated with an equally weighted binning of all limit orders. In such a binning, orders with short lifetimes dominate due to their large number, although they contribute almost nothing to filling the limit order book with volume. Therefore we propose to weight the binning with lifetimes of limit orders. To analyze the effect of a weighted binning as compared to an equal binning for all limit orders, the limit order volume is the best quantity to start with. As above, we use the inverse cumulative distribution function (Eq.~\ref{eq:icdf}) and find that it is broad as well for the limit order volume, as presented in Fig.~\ref{fig:weighted_volume}. Fitting it with a power law, as in \cite{Maslov.2001},
\begin{align}
P(v_i) \sim& \frac{1}{v_i^{2.4}} ,
\end{align}
emphasizes how broad the volumes are distributed, even though they are not really power law distributed. Volumes are strongly concentrated at round values as 100 shares or 5000 shares, where the icdf has jumps. Similar observations were made in \cite{Challet.2001, Mu.2009}. Weighting the contribution of each limit order to the volume distribution with its lifetime, the icdf is strongly increased, especially for large volumes between $10^3$ shares and $10^4$ shares by more than an order of magnitude. 

\begin{figure}[htb]
	\begin{center}
		\includegraphics[width=0.98\columnwidth]{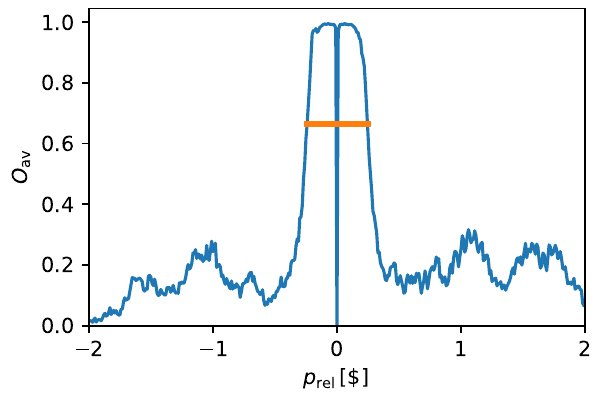}
		\includegraphics[width=0.98\columnwidth]{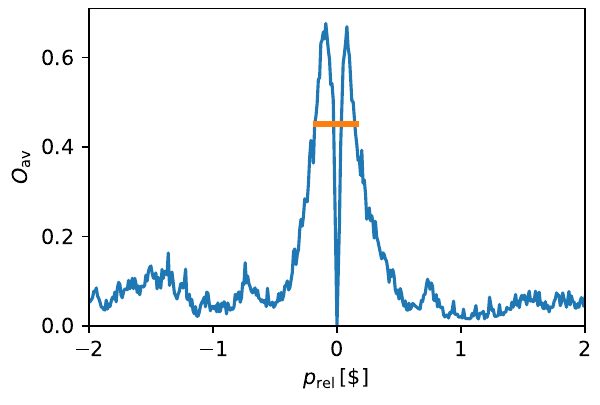}
		\caption{Top: Time averaged occupation of relative price levels for \textsc{Ebay} on March 10, 2016. The orange horizontal lines indicate the full width at two-thirds of the maximum. Bottom: Same for \textsc{Liberty Interactive Corporation}}
		\label{fig:average_occupation}
	\end{center}
\end{figure}

Effects of weighting with lifetimes are even more pronounced for the distribution of relative insertion prices defined as
\begin{align}
p_{{\rm rel},i} &= \begin{cases} \displaystyle{\frac{p_i-m(t_i)}{m(t_i)}} & {\rm for\,sell\,limit\,orders}\\
\displaystyle{\frac{m(t_i)-p_i}{m(t_i)}} & {\rm for\,buy\,limit\,orders}
\end{cases}
\end{align}
with the insertion price $p_i$ of the limit order and the midpoint price immediately before insertion $m(t_i)$. Relative insertion prices can be negative, if the inserted order crosses the midpoint price. In such a case the quote changes and the spread reduces. For example, if we have a quote of $0.98\,\$$ (bestbid) and $1.02\,\$$ (bestask) the midpoint price is $1.00\,\$$. If we now insert a buy limit order at $1.01\,\$$ the relative insertion price is $p_{{\rm rel},i} = -0.01$. Our new quote is $1.01\,\$$ (bestbid) and $1.02\,\$$ (bestask) and the spread reduces from $0.04\,\$$ to $0.01\,\$$. For buy limit orders, relative insertion prices are always smaller than one, for sell limit orders they can be larger (for $p_i$ larger than $2 m(t_i)$). The differences between weighted and unweighted icdfs of relative insertion prices in Fig.~\ref{fig:weighted_price} (combined for buy and sell limit orders) are huge. For example, insertions with relative prices $p_{{\rm rel},i}>10^{-2}$ include only 3\,\textperthousand \ of all limit orders, but such orders contribute almost half of the limit order book filling, with a lifetime-weighted cumulative share of 45\,\%. Weighting with the volume instead of the lifetime has only small effects. Weighting with the product of lifetime and volume is the most significant measure with respect to the total volume stored in the limit order book. The icdf shown in Fig.~\ref{fig:weighted_price} does not reach a value of one because limit order insertions hitting or undercutting the midpoint price $m(t_i)$ are not included due to the double-logarithmic scale. Limit order insertions hit or undercut the midpoint price $m(t_i)$ in 30\,\% of all cases. Such limit orders are mostly very short-living, hence the effect is only weakly stressed for a lifetime weighted icdf. As a measure of the average cost of the stock we introduce the daily average midpoint price $\overline{m}$. With a daily average midpoint price of $\overline{m}=23.4\,\$$, the smallest positive relative insertion prices are $p_{{\rm rel},i}\approx 2.1\cdot 10^{-4}$ (half cent from midpoint) and $p_{{\rm rel},i}\approx 4.2\cdot 10^{-4}$ (one cent from midpoint). Especially at the latter there is a large jump in the icdf. This is in accordance with \cite{Challet.2001, Bouchaud.2002, Gu.2008}, where order insertions were found to be close to the quotes.

\begin{figure}[htb]
	\begin{center}
		\includegraphics[width=0.98\columnwidth]{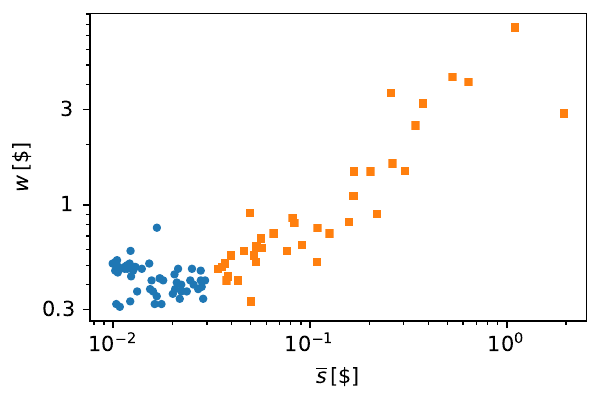}
		\caption{Scatter plot of cushion width over time averaged spread on a double-logarithmic scale. Large-tick stocks with small spread of $\overline{s}\leq 3$ cents are shown with blue circles, small-tick stocks with orange squares.}
		\label{fig:v_vs_spread}
	\end{center}
\end{figure}

\begin{figure}[htb]
	\begin{center}
		\includegraphics[width=0.98\columnwidth]{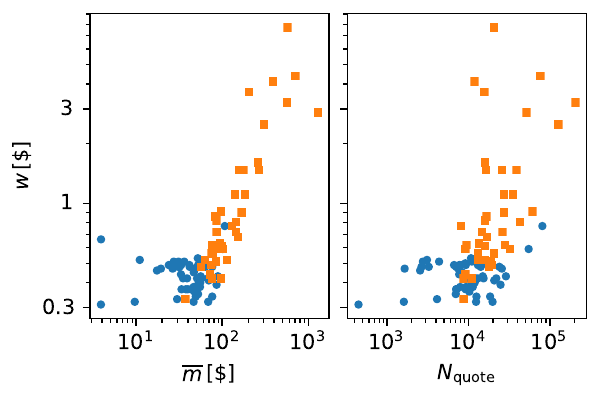}
		\caption{Left: Scatter plot of cushion width over time averaged midpoint price on a double-logarithmic scale. Right: Scatter plot of cushion width over number of daily quote changes on a double-logarithmic scale. Large-tick stocks with small spread of $\overline{s}\leq 3$ cents are shown with blue circles, small-tick stocks with orange squares. }
		\label{fig:v_vs_price}
	\end{center}
\end{figure}

\subsection{Width of the liquidity cushion}
\label{Cushion}

\begin{figure}[htb]
	\begin{center}
		\includegraphics[width=0.98\columnwidth]{figures/cushion_width_analysis_Vcap.pdf}
		\caption{Scatter plot of cushion width over total amount of money traded per trading day on a double-logarithmic scale. Large-tick stocks with small spread of $\overline{s}\leq 3$ cents are shown with blue circles, small-tick stocks with orange squares.}
		\label{fig:v_vs_Vcap}
	\end{center}
\end{figure}

\begin{figure}[htb]
	\begin{center}
		\includegraphics[width=0.98\columnwidth]{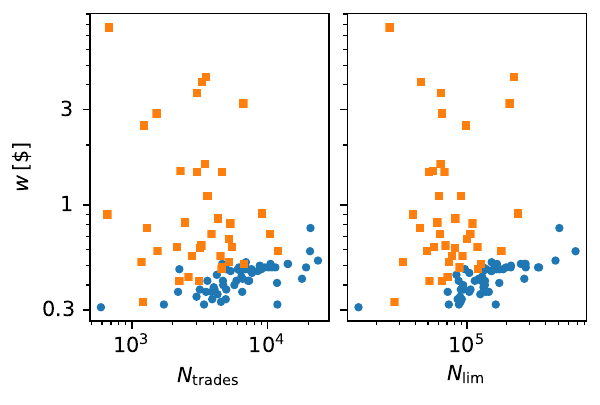}
		\caption{Left: Scatter plot of cushion width over average number of trades per trading day on a double-logarithmic scale. Right: Scatter plot of cushion width over average number of limit orders per trading day on a double-logarithmic scale. Large-tick stocks with small spread of $\overline{s}\leq 3$ cents are shown with blue circles, small-tick stocks with orange squares.}
		\label{fig:v_vs_N}
	\end{center}
\end{figure}

\begin{figure}[htb]
	\begin{center}
		\includegraphics[width=0.98\columnwidth]{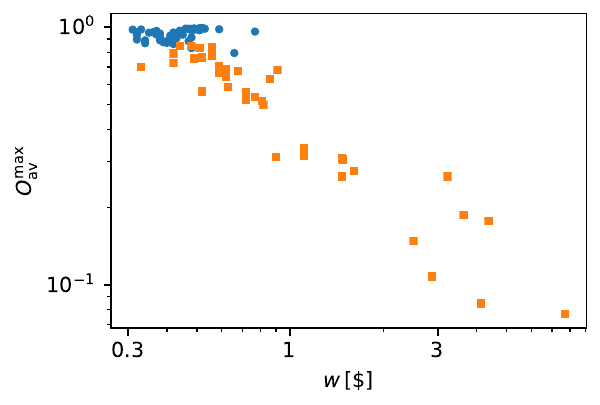}
		\caption{Scatter plot of maximum time averaged occupation over cushion width on a double-logarithmic scale. Large-tick stocks with small spread of $\overline{s}\leq 3$ cents are shown with blue circles, small-tick stocks with orange squares.}
		\label{fig:v_vs_h}
	\end{center}
\end{figure}

We introduce a method for identifying the width of 
the densely filled regime around the midpoint price which exists for many stocks in a similar fashion as shown in Fig.~\ref{fig:LOB_Ebay} and Fig.~\ref{fig:LOB_LVNTA}. We call this regime the liquidity cushion. As this regime moves with the midpoint price, we introduce the relative price 
\begin{align}
p_{\rm rel}(t) &= p - m(t),
\end{align}
where $p$ is a price niveau of the limit order book. Note that we are not dividing by the midpoint price here. The limit order book volume expressed with the relative price is defined as
\begin{align}
v_{\rm rel}(p_{\rm rel},t) &= v(p_{\rm rel} + m(t),t).
\end{align}
Long-living orders can contribute to volumes at varying relative prices due to changes of the midpoint price $m(t)$. As $v_{\rm rel}$ contains large volumes even far from the current midpoint price, the densely filled regime around the midpoint price is camouflaged. Therefore we use a different observable distinguishing the densely filled regime and the fragmented distant field, the time averaged occupation 
\begin{align}
O_{\rm av}(p_{\rm rel}) &= \langle \Theta(v_{\rm rel}(p_{\rm rel},t))\rangle_t
\end{align}
with the Heaviside function
\begin{align}
	\Theta(v) &= \begin{cases} 0 & \textrm{for } v = 0 \\
		1 & \textrm{for } v > 0
	\end{cases}
\end{align}
and $\langle . \rangle_t$ being the time average over the trading hours between 9:30 and 16:00. The time averaged occupation is in the regime \mbox{$0\leq O_{\rm av}(p_{\rm rel})\leq 1$}. The limiting cases are $O_{\rm av}(p_{\rm rel})=0$ if the volume at $p_{\rm rel}$ is empty during the whole time and $O_{\rm av}(p_{\rm rel})=1$ if it is occupied during the whole time. The time averaged occupation for \textsc{Ebay} reaches values close to one around a relative price of zero as shown in Fig.~\ref{fig:average_occupation}, top. The occupation at exactly $p_{\rm rel}=0$ is zero, because it is impossible for a limit order to be located at the midpoint price. The occupation drops to smaller values for relative prices far from zero. As a measure for the average width $w$ of the densely filled regime around the midpoint price we use the full width at two-thirds of the maximum because the occupation saturates at values above zero outside the peak we want to characterize. We identify  $w=49$ cents as average width of the densely filled regime around the midpoint price for \textsc{Ebay} in Fig.~\ref{fig:LOB_Ebay}. We can transform $w$ into $w_{\rm rel} = w/2\overline{m}\approx 0.01$ and compare this to the relative insertion prices. In Fig.~\ref{fig:weighted_price} we see that the inverse cumulative distribution of relative insertion prices drops strongly before this value is reached, such that most limit orders are inserted in the liquidity cushion with a relative insertion price below this value. However, the distribution of relative insertion prices weighted with lifetime in Fig.~\ref{fig:weighted_price} shows that the few orders inserted with larger relative insertion prices still contribute a lot to filling the limit order book. 

The time averaged occupation for \textsc{Liberty Interactive Corporation} shown in Fig.~\ref{fig:average_occupation}, bottom, is similar besides the maximum not being in a regime around $p_{\rm rel}=0$ but a little farther away because for this small-tick stock the spread is often larger than one tick. We find a full width at two-thirds of the maximum of $w=36$ cents which seems reasonable with Fig.~\ref{fig:LOB_LVNTA}. 

\begin{figure}[htb]
	\begin{center}
		\includegraphics[width=0.98\columnwidth]{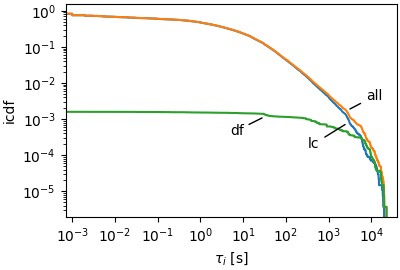}
		\includegraphics[width=0.98\columnwidth]{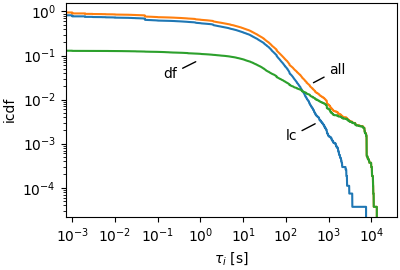}
		\caption{Top: Inverse cumulative distribution function of lifetimes of limit orders during the market hours between 9:30 and 16:00 (orange line, `all') on a double-logarithmic scale for \textsc{Ebay} on March 10, 2016. Contribution of limit orders introduced in the liquidity cushion (blue line, `lc') and in the fragmented distant field (green line, `df'). Bottom: Same for \textsc{Liberty Interactive Corporation}.}
		\label{fig:lifetime_regimes}
	\end{center}
\end{figure}

\begin{figure}[htb]
	\begin{center}
		\includegraphics[width=0.98\columnwidth]{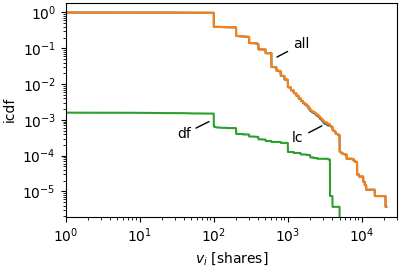}
		\includegraphics[width=0.98\columnwidth]{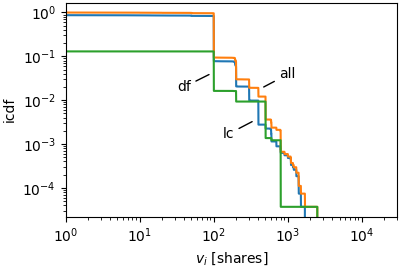}
		\caption{Top: Inverse cumulative distribution function of limit order volumes during the market hours between 9:30 and 16:00 (orange line, `all') on a double-logarithmic scale for \textsc{Ebay} on March 10, 2016. Contribution of limit orders introduced in the liquidity cushion (blue line, `lc') and in the fragmented distant field (green line, `df'). Bottom: Same for \textsc{Liberty Interactive Corporation}.}
		\label{fig:volume_regimes}
	\end{center}
\end{figure}

\subsection{Relationship between width of the liquidity cushion and other trading characteristics}
\label{CushionComp}

Based on the time averaged occupation over five trading days we calculate the cushion width $w$ for all stocks and analyze how it is influenced by different order flow and price characteristics. In Fig.~\ref{fig:v_vs_spread} every data point represents the cushion width $w$ in dependence of the time averaged spread $\overline{s}$ over five trading days for one stock. In the following we distinguish between large-tick and small-tick stocks, with large-tick stocks having an time averaged spread $\overline{s}<3$ cents. For a small spread the cushion width saturates above 30 cents. For large-tick stocks the cushion width even decreases slightly with the spread size. Cushion widths increase considerably with the spread size for small-tick stocks with a large spread, with largest values more than ten times larger than the smallest cushion width. A similar dependence structure can be seen for the time averaged midpoint price $\overline{m}$ in Fig.~\ref{fig:v_vs_price} on the left and for the average number of quote changes $N_{\rm quote}$ per trading day on the right. The cushion width increases with the average total amount of money involved in all trades per trading day, $V_{\rm cap}$. This holds for all stocks, as can be seen in Fig.~\ref{fig:v_vs_Vcap}. It is also possible that only large-tick stocks show a dependence on a quantity while small-tick stocks are not correlated with it. This is the case for the average number of trades $N_{\rm trades}$ respectively the average number of limit orders $N_{\rm lim}$ per trading day, shown in Fig.~\ref{fig:v_vs_N}. Finally, for large-tick and small-tick stocks cushion widths are not correlated with the volatility of stock prices. 

\begin{figure*}[htb]
	\begin{center}
		\includegraphics[trim=0 0 30 0,clip,width=0.98\textwidth]{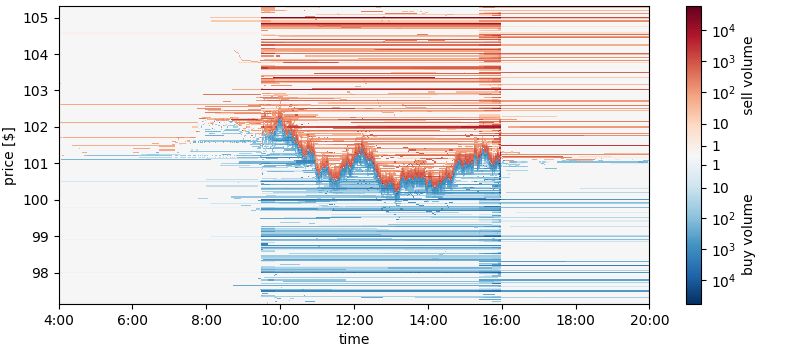}
		\includegraphics[width=0.98\columnwidth]{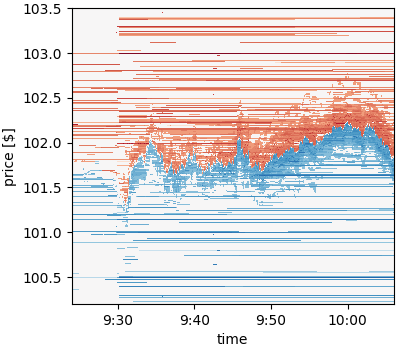}
		\includegraphics[width=0.98\columnwidth]{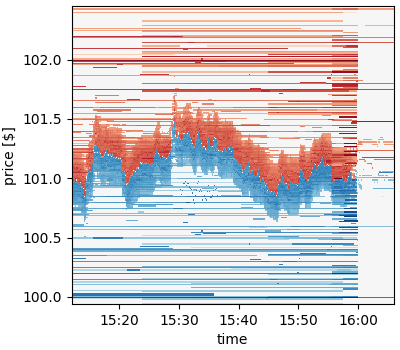}
		\caption{Top: Color plot of the time evolution of volume in the limit order book over a section of limit order prices around the midpoint price. Volumes are shown on a logarithmic color scale (red for sell limit orders and blue for buy limit orders) for \textsc{Apple} on March 10, 2016. Bottom left: Detailed view of the market opening. Bottom right: Detailed view of the market closing.}
		\label{fig:LOB_Apple}
	\end{center}
\end{figure*}

So far we found that the cushion width is influenced by the stock price and many order flow characteristics. Next we analyze how on the other hand the cushion width influences the limit order book appearance. Fig.~\ref{fig:v_vs_h} shows that the maximum time averaged occupation
\begin{align}
	O_{\rm av}^{\rm max} &= \max_{p_{\rm rel}} O_{\rm av}(p_{\rm rel})
\end{align}
is almost one for large-tick stocks as in Fig.~\ref{fig:LOB_Ebay} and Fig.~\ref{fig:average_occupation}, top, meaning that gaps are only rarely present in the liquidity cushion. For small-tick stocks as in Fig.~\ref{fig:LOB_LVNTA} and Fig.~\ref{fig:average_occupation}, bottom, $O_{\rm av}^{\rm max}$ is smaller and decreases with the cushion width. Hence for large cushion widths there are large gaps between occupied levels in the liquidity cushion.

\subsection{Characterizing liquidity cushion and fragmented distant field}
\label{distant field}

\begin{figure*}[htb]
	\begin{center}
		\includegraphics[width=0.98\textwidth]{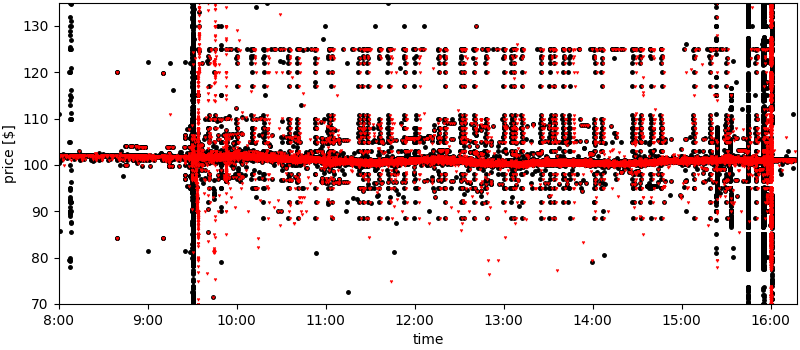}
		\caption{Scatter plot of limit order insertion prices versus times (black circles) and cancellation prices versus times (red triangles) for \textsc{Apple} on March 10, 2016.}
		\label{fig:Insertions_Apple}
	\end{center}
\end{figure*}

\begin{figure}[htb]
	\begin{center}
		\includegraphics[width=0.98\columnwidth]{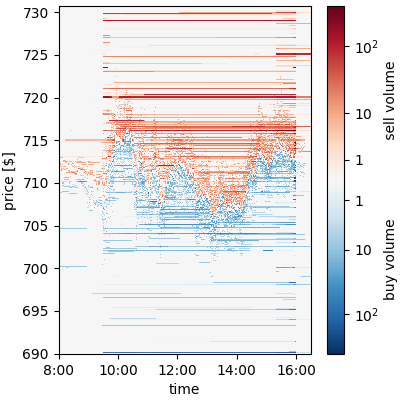}
		\caption{Color plot of the time evolution of volume in the limit order book over a section of limit order prices around the midpoint price. Volumes are shown on a logarithmic color scale (red for sell limit orders and blue for buy limit orders) for \textsc{Google} on March 10, 2016.}
		\label{fig:LOB_Google}
	\end{center}
\end{figure}

\begin{figure}[htb]
	\begin{center}
		\includegraphics[width=0.98\columnwidth]{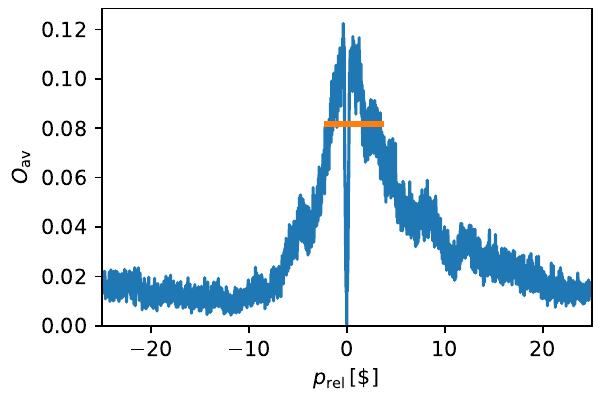}
		\caption{Time averaged occupation of relative price levels for \textsc{Google} on March 10, 2016. The orange horizontal line indicates the full width at two-thirds of the maximum.}
		\label{fig:average_occupation_Google}
	\end{center}
\end{figure}

\begin{figure*}[htb]
	\begin{center}
		\includegraphics[width=0.98\textwidth]{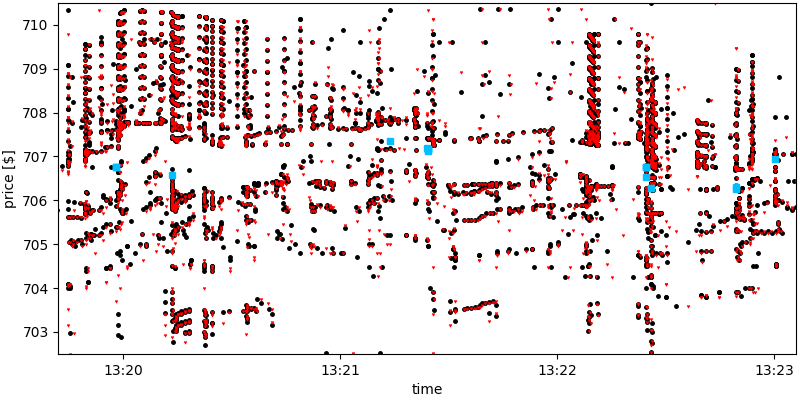}
		\caption{Scatter plot of limit order insertion prices versus times (black circles), cancellation prices versus times (red triangles) and trade prices versus times (blue squares) for \textsc{Google} in a short time interval on March 10, 2016.}
		\label{fig:Insertions_Google}
	\end{center}
\end{figure*}

The characteristics of orders inserted in the liquidity cushion ($p_{\text{rel}, i} m(t_i)\leq w/2$) are very different than those of orders inserted in the fragmented distant field ($p_{\text{rel}, i} m(t_i)>w/2$). The share of the inverse cumulative lifetime distribution for limit orders inserted in the fragmented distant field is concentrated on large lifetimes as can be seen in Fig.~\ref{fig:lifetime_regimes}. As this share is normalized to the total number of limit orders during the trading hours, it reaches a plateau far below one for small lifetimes, corresponding to the share of limit orders inserted in the fragmented distant field among all limit orders of $0.0016$ for \textsc{Ebay} and $0.12$ for \textsc{Liberty Interactive Corporation}. 
The contribution of orders inserted in the liquidity cushion dominates the whole lifetime distribution, with strong deviations only for long lifetimes. The lifetime distribution is dominated by limit orders inserted in the fragmented distant field for longest lifetimes. Such orders make a large contribution to filling the limit order book. Therefore, the distribution of relative prices weighted with lifetime in Fig.~\ref{fig:weighted_price} keeps being at large levels even in the fragmented distant field which is above $p_{{\rm rel},i}=0.01$ for this particular case. This can also be understood with the average lifetime of orders inserted in the liquidity cushion of 32\,s compared to a much larger average lifetime of orders inserted in the fragmented distant field of 2488\,s. The total lifetime distribution for Liberty Interactive Corporation has a double-humped structure. We see similar results for many of the 96 stocks under consideration. The contribution of the liquidity cushion is consistent with a power law with an exponent between one and two for the icdf. This fits with findings in \cite{Eisler.2009}, where the power law of lifetimes is understood as a mix of order cancellations and trades. The distribution of lifetimes in the distant field has a much broader tail.

For the order volumes the inverse cumulative distribution is also dominated by orders inserted in the liquidity cushion as can be seen in Fig.~\ref{fig:volume_regimes}. For \textsc{Ebay} the total distribution and the share due to limit orders inserted in the liquidity cushion are almost completely on top of each other. The share in the fragmented distant field is concentrated at medium sized volumes and the largest volumes are not present. This implies the average volume of orders inserted in the liquidity cushion of 205 shares per limit order is in the same order of magnitude as the average volume of orders inserted in the fragmented distant field of 436 shares per limit order. Therefore the inverse cumulative distribution of relative insertion prices changes in the fragmented distant field ($p_{{\rm rel},i}>0.01$) by a factor of two at most when weighted with the volumes, as seen in Fig.~\ref{fig:weighted_price}. 
The contributions to the inverse cumulative distribution of volumes for \textsc{Liberty Interactive Corporation} (Fig.~\ref{fig:volume_regimes}, bottom) behave in a similar way as for \textsc{Ebay}, but here the contribution of limit orders inserted in the fragmented distant field is much more significant and even dominates the whole distribution for medium sized volumes around 500 shares per limit order.

\begin{figure}[htb]
	\begin{center}
		\includegraphics[width=0.98\columnwidth]{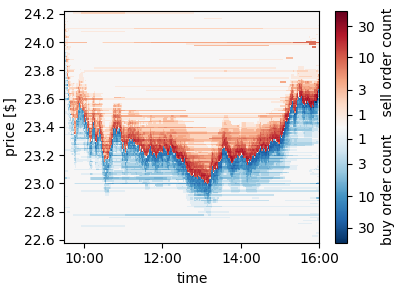}
		\caption{Time-dependent sell and buy order count for \textsc{Ebay} on March 10, 2016.}
		\label{fig:LOB_count_Ebay}
	\end{center}
\end{figure}

The fragmented distant field is particularly rich for the highly traded \textsc{Apple} stock, as can be seen in Fig.~\ref{fig:LOB_Apple}. The overview of the whole market day on the top is realized by binning prices together in groups of three cents and dividing the volume by three in order to show the average volume per price level. This is necessary because the relevant price regime spans more than 800 cents, which is too much for the figure resolution. Pre-market hours start at 4:00. In the early morning hours only a few limit orders with small volume are inserted. Stronger activity starts around 8:00, but still the liquidity cushion does not form. Directly after the market opening at 9:30 the limit order book is deeply filled. The detailed view of the opening, shown in Fig.~\ref{fig:LOB_Apple}, bottom left, reveals that the fragmented distant field is filled in a short time period of about a minute after market opening. For the detailed view the binning of price levels is not necessary, as the price regime covered is smaller. A certain fraction of the limit orders in the fragmented distant field is cancelled within the first half hour after market opening. The fragmented distant field is filled more strongly during the first period after market opening than later on. The filling of the fragmented distant field could be a reaction to a larger volatility and a larger probability of flash crashes during the time after market opening to either utilize or depress them. On the other hand, the liquidity cushion builds up only slowly. It needs about half an hour to obtain its later form mostly without any gaps. During the trading hours between 9:30 and 16:00, many limit orders in the fragmented distant field are long living and adaptations to current trades and quotes are of minor importance. On the other hand the liquidity cushion closely follows the midpoint price with a symmetric appearance on sell and buy side and even jumps of the midpoint price are followed closely. As shown in Fig.~\ref{fig:LOB_Apple}, bottom right, in the last hour before market closing again a large number of extra limit orders is inserted into the fragmented distant field, possibly again as a reaction to flash crashes being more likely shortly before the market closes. This happens in a concentrated way around 15:23, 15:45 and a few minutes before market closing. After market closing at 16:00 the liquidity cushion vanishes immediately and the fragmented distant field is sparse. 

\begin{figure}[htb]
	\begin{center}
		\includegraphics[width=0.98\columnwidth]{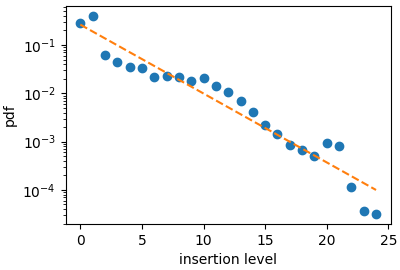}
		\caption{Frequencies of limit order insertions depending on the insertion level on a logarithmic scale (dots) with an exponential fit (dashed line).}
		\label{fig:level_Ebay}
	\end{center}
\end{figure}

By looking at individual insertions and cancellations of limit orders over the trading day, as shown in Fig.~\ref{fig:Insertions_Apple}, we can learn more about characteristic times and prices of the fragmented distant field. Every black dot marks the price and the time of a limit order insertion, every red dot marks the price and time of a limit order cancellation. Black dots form vertical lines which means that many limit orders are inserted in a short period of time at different prices. This happens shortly after eight, directly after market opening at 9:30, several times in the last hour before market closing, and shortly after market closing. Limit order cancellations are clustered in time as well, especially a few minutes after market opening and at market closing. At certain round prices alternating insertions and cancellations additionally form horizontal lines, large numbers of limit orders are inserted and cancelled thus constantly adapted during the whole trading day. This happens even far from the current price. Many other price levels in the fragmented distant field are occupied by limit orders which stay constant during the whole trading hours, with one black point at 9:30, one red point at 16:00 and nothing in between. 

The behavior of small-tick stocks is similar, with a static fragmented distant field and a liquidity cushion that is adapting fast to midpoint price changes. In Fig.~\ref{fig:LOB_Google} the time dependent limit order book of \textsc{Google} was realized by binning prices together in groups of twelve cents. This is again necessary because the relevant price regime spans 4000 cents, which is too much for the figure resolution. The volume is in general much smaller than for all stocks discussed so far, because the stock price is much larger. The average volume is only 70 shares per limit order, 55\,\% of all limit orders have a volume less than 100 shares. 
Further we see that liquidity cushion and static fragmented distant field overlap with only little interaction. The liquidity cushion starts to build up already before the market opening. The time averaged occupation shown in Fig.~\ref{fig:average_occupation_Google} implies a cushion width $w=5.43\,\$ $ together with a maximum occupation of $O_{\rm av}^{\rm max} = 0.123$. The latter is consistent with large gaps in the liquidity cushion. Averaging over five trading days the cushion width is $w=4.37\,\$ $.

In Fig.~\ref{fig:Insertions_Google} limit order insertions, cancellations and trades for a regime around the liquidity cushion are presented for a short time window. There are much more limit order insertions and cancellations than trades. There are sequences of insertions and cancellations with slightly changing prices in short time intervals which form short curved lines. Some of these structures exist at the same time at many different prices synchronously. Such a rich structure is rare, at least among the 96 stocks we analyze here. This is because for stocks with a smaller cushion width $w$ a large number of insertions and cancellations is concentrated on a few price levels. Most likely, similar structures exist but they are overlapping so much that they cannot be recognized easily. 

\FloatBarrier
\section{Modeling the liquidity cushion}
\label{Model}

\begin{figure}[htb]
	\begin{center}
		\includegraphics[width=0.98\columnwidth]{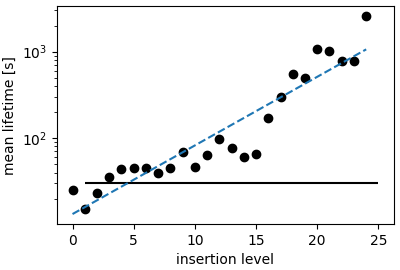}
		\caption{Average limit order lifetime depending on the insertion level on a logarithmic scale (dots) with an exponential fit (dashed line) and average lifetime of all limit orders (horizontal solid line).}
		\label{fig:level_lifetime_Ebay}
	\end{center}
\end{figure}

\begin{figure*}[htb]
	\begin{center}
		\includegraphics[width=0.98\textwidth]{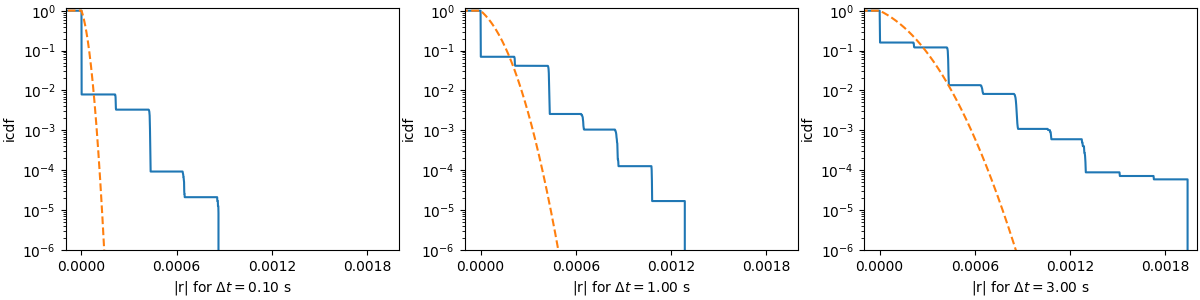}
		\includegraphics[width=0.98\textwidth]{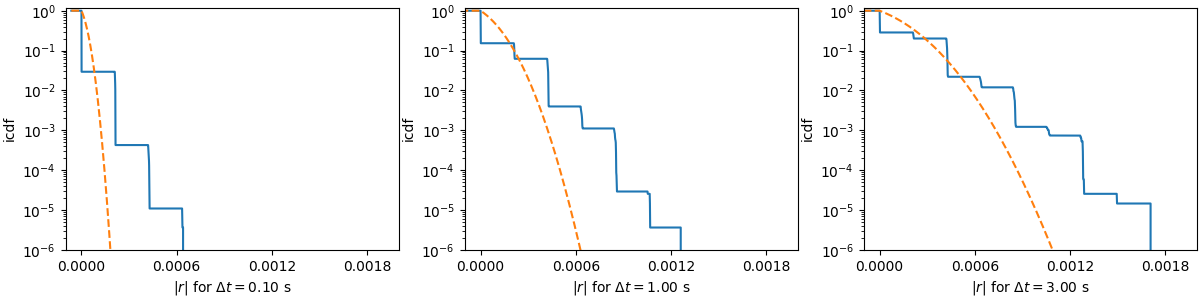}
		\caption{Top: Inverse cumulative distribution function of the absolute return (blue line) with a Gaussian fit (orange dashed line) for different time spans for \textsc{Ebay}. Bottom: Same for the model.}
		\label{fig:ret_dist_Ebay_ABM}
	\end{center}
\end{figure*}

After what we learned from Figs.~\ref{fig:LOB_Apple}-\ref{fig:LOB_Google}, the liquidity cushion and the fragmented distant field need to be modelled in completely different ways. In the fragmented distant field there is not much of interaction of limit orders among each other and with the quotes. Therefore, it would be easy to construct a model based on the power law distributed relative price and characteristic times as opening, a period after opening prone to flash crashes, the market closing etc.

We design a \textit{zero intelligence} model for the liquidity cushion with stochastic insertion of limit orders and market orders, in the style of \cite{Cont.2010, Bouchaud.2002, Eisler.2009, Maslov.2000, Toth.2009}. Ignoring the volume distribution and setting all limit order volumes to the average volume, we are left with modeling the limit order count. We concentrate on large-tick stocks and particularly on \textsc{Ebay}. The according time-dependent sell and buy order count can be seen in Fig.~\ref{fig:LOB_count_Ebay}. The maximal order count exceeds thirty and is largest close to the quotes. 

\begin{figure}[htb]
	\begin{center}
		\includegraphics[width=0.98\columnwidth]{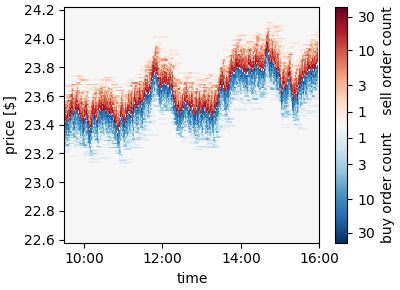}
		\caption{Time-dependent sell and buy order count for the model.}
		\label{fig:LOB_count_ABM}
	\end{center}
\end{figure}

\begin{figure}[htb]
	\begin{center}
		\includegraphics[width=0.98\columnwidth]{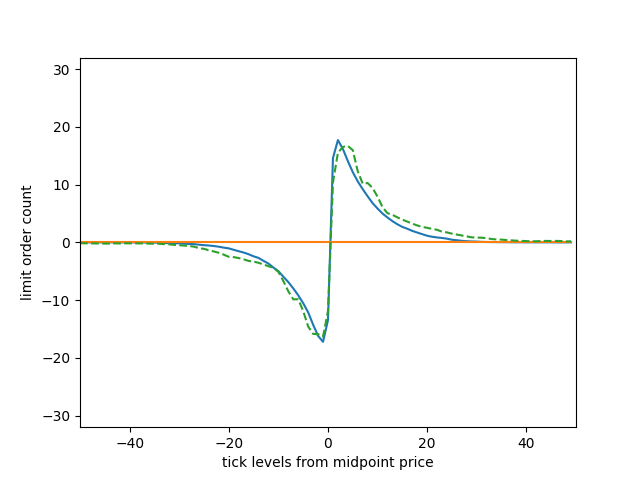}
		\caption{Average limit order book filling for the model (blue line) in comparison to the average limit order book filling for \textsc{Ebay} (green dashed line).}
		\label{fig:avg_LOB_Ebay_ABM}
	\end{center}
\end{figure}

\begin{figure}[htb]
	\begin{center}
		\includegraphics[width=0.6\columnwidth]{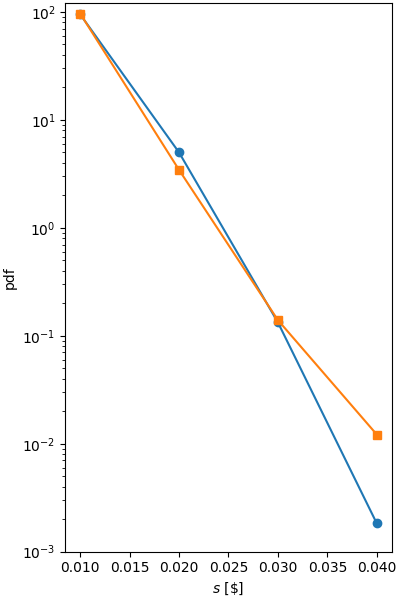}
		\caption{Frequencies of spread widths on a logarithmic scale for \textsc{Ebay} (orange squares) and the model (blue circles).}
		\label{fig:spread_Ebay_ABM}
	\end{center}
\end{figure}

\begin{figure}[htb]
	\begin{center}
		\includegraphics[width=0.98\columnwidth]{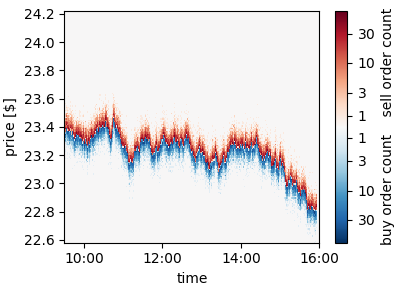}
		\includegraphics[width=0.98\columnwidth]{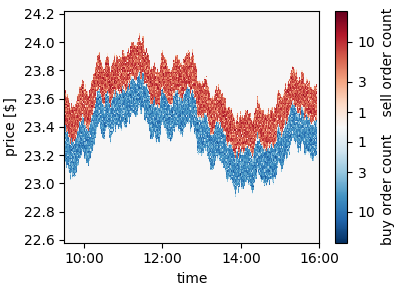}
		\caption{Top: Time-dependent sell and buy order count for the simplified model with uniform lifetimes. Bottom: Same for the simplified model with uniform insertion probabilities and lifetimes.}
		\label{fig:LOB_count_ABM_variants}
	\end{center}
\end{figure}

\begin{figure}[htb]
	\begin{center}
		\includegraphics[width=0.98\columnwidth]{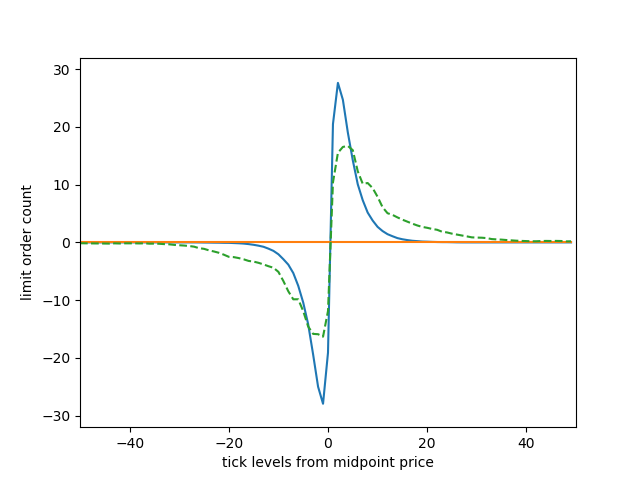}
		\includegraphics[width=0.98\columnwidth]{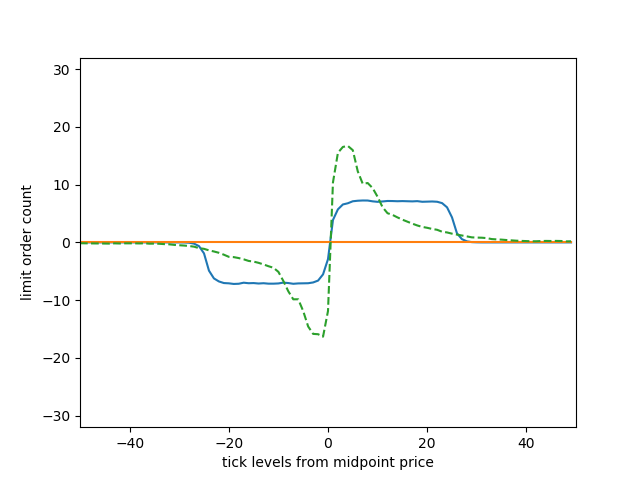}
		\caption{Top: Average limit order book filling for the simplified model with uniform lifetimes (blue line) in comparison to the average limit order book filling for \textsc{Ebay} (green dashed line). Bottom: Same for the simplified model with uniform insertion probabilities and lifetimes.}
		\label{fig:avg_LOB_Ebay_ABM_variants}
	\end{center}
\end{figure}

We define the model dynamics with sequential insertion of $N$ orders during one trading day of duration $T$ with identical time distances in between order insertions of $T/N$. In the beginning, the limit order book is filled around the start price $S_0$ with ten limit orders on each tick up to $S_0\pm L$ with parameter $L$ indicating the half width of the liquidity cushion. Initial limit orders have a lifetime of 30\,s. The properties of all further orders are assigned in the following random fashion.  
\begin{itemize}
\item With probabilities of 1/2, an order is sell or buy. 
\item With probability $P_{\rm market}$ the order is a market order, otherwise it is a limit order. 
\item For a buy (sell) market order the volume is adjusted exactly to the volume on the best sell (buy) price in the limit order book. 
\item For determining the parameters of a limit order an integer level $0\leq l< L$ is chosen with probability 
\begin{align}
P_l=c \exp(-l/l_0), \quad \sum\limits_{l=0}^{L-1}P_l=1
\end{align}
with normalization constant $c$ and the model parameter of typical level depth $l_0$.
\item The insertion price for a buy (sell) limit order is set to $l+1$ ticks below (above) the best sell (buy) price. 
\item The lifetime of the limit order is set to $t_{\rm lt} \exp(l/l_{\rm lt})$. After this time the limit order is cancelled, if it was not traded already. The shortest lifetime $t_{\rm lt}$ is assigned at level $l=0$. 
\end{itemize}

The model parameters $N$, $T$, $L$, $P_{\rm market}, l_0$, $t_{\rm lt}$ and $l_{\rm lt}$ are adjusted to the order flow of \textsc{Ebay} on March 10, 2016. The trading hours last six and a half hours, hence we have $T=23400\,$s. For the width of the liquidity cushion we found above $w=49$ cents. Accordingly we set $L=25$. From 13429 trades we reconstruct 4035 market orders. Almost two thirds of these market orders empty the best buy or best sell offer completely, most of the remaining market orders empty a large part of the best offer and are traded against many limit orders. Therefore, it is reasonable to model market orders as being adjusted to the volume of the respective best offer. The number of all limit orders is 269800, so the total order count adds up to $N=273835$. The probability of an order being a market order is set to the share of market orders among all orders, $P_{\rm market}=0.0147$. The remaining parameters are obtained from exponential fits. For insertion probabilities depending on insertion level we find a strongly decreasing function over four orders of magnitude, shown in Fig.~\ref{fig:level_Ebay}. Performing a linear fit to the logarithmic data we find $l_0=3.045$. In this way  relative errors are minimized, while by performing an exponential fit to the original data the absolute errors would be minimized. The broad distribution of lifetimes is reflected in strongly increasing average lifetimes of limit orders with the insertion level over two orders of magnitude, as seen in Fig.~\ref{fig:level_lifetime_Ebay}. Performing a linear fit to the logarithmic data we find $t_{\rm lt}=13.24\,$s and $l_{\rm lt}=5.46$. The average lifetime of all limit orders inserted in the liquidity cushion is $30.019\,$s.

Our model is similar to the model in \cite{Bouchaud.2002}. The most important difference is that the limit order lifetimes depend on the relative insertion price in the model presented here. The distribution of relative prices (the level $l$) is modeled by an exponential here, as the power law dependence holds best for large distances from the quotes, and for prices a few ticks from the quote we find the exponential fit to be better. In \cite{Bouchaud.2002} the empirical distribution is used directly without fit and the deep filling of the order book is included in the model, while we restrict the model to the liquidity cushion. The model in \cite{Eisler.2009} is more complicated than our model, as it considers an individually adapted power law distribution of cancellation times for every insertion level. This implies many more parameters than in our model and is not needed for an adequate modeling of the order book characteristics, as we see in the following.

To validate the adequacy of our model we use the following characteristics: return distribution, (daily) price range, time-dependent limit order book, average limit order book filling and spread distribution.
We define the return from a time $t$ to a later time $t+\Delta t$ as
\begin{align}
	r(t,\Delta t) = \ln\frac{m(t+\Delta t)}{m(t)} .
\end{align}
The distribution of absolute returns for $\Delta t =$ 0.10\,s, 1.00\,s and 3.00\,s for the model and the \textsc{Ebay} stock are shown in Fig.~\ref{fig:ret_dist_Ebay_ABM}. For all time spans the empirical distribution and the model distribution show much heavier tails than the respective normal distribution fitted to the standard deviation. Comparing the distributions for the model and \textsc{EBAY}, they differ strongly for the shorter time span of $\Delta t =$ 0.10\,s. For the longer time span of $\Delta t =$ 1.00\,s, the model distribution fits the empirical distribution and for the time span of 3.00\,s, deviations only occur for large returns $\lvert r \rvert >$ 0.0012. Comparing the time-dependent order count for the model and empirical data in Fig.~\ref{fig:LOB_count_Ebay} and Fig.~\ref{fig:LOB_count_ABM}, we find not only the price range but also the optical appearance to be similar to each other. Price movement and jumps are followed closely by the liquidity cushion. The order count decreases with the distance to the quotes and the solid inner liquidity cushion gets more and more blurred and smooth. The average limit order book filling, as seen in Fig.~\ref{fig:avg_LOB_Ebay_ABM}, supports the impression. The average number of limit orders per level shows a nearly identical course for the model and the empirical data. Only price levels far away from the quotes are slightly underrepresented in the model. This could at least partly be caused by the contribution of the fragmented distant field to the empirical distribution. Finally the spread distribution shows for both cases an exponential decay over nearly four orders of magnitude from the smallest possible spread of 0.01\,\$ to a spread of 0.04\,\$, as seen in Fig.~\ref{fig:spread_Ebay_ABM}.

We establish two additional simplified model versions which help to understand the implications of inhomogeneous insertion probabilities and broadly distributed lifetimes. For a simplified model with uniform lifetimes, we assign the average lifetime of $30.019\,$s to each order, irrespective of the level where it is inserted. For a further simplified model with a uniform insertion probability, we set $P_l=1/L$ (corresponding to $l_0\to \infty$) and additionally assign the average lifetime of $30.019\,$s to each order.
For the simplified models we do not find notable differences for the return distribution, price range and even the spread distribution. By contrast we see strong distinctions for the optical appearance of the time-dependent order count, shown in Fig.~\ref{fig:LOB_count_ABM_variants}, and for the average limit order book filling, shown in Fig.~\ref{fig:avg_LOB_Ebay_ABM_variants}. Neglecting the different average lifetimes of the different insertions levels, we end up with a much narrower, solid liquidity cushion. Levels close to the quotes are highly overrepresented and those farther away highly underrepresented. Neglecting also the different insertion probabilities of the insertion levels, we end up with an almost uniform distribution of limit orders in the liquidity cushion with a much lower maximum order count. These results emphazise the importance of the distribution of limit order lifetimes and insertion probabilities for modeling the average filling of the limit order book.

In neither of our models we allow for adaptions of limit orders after they are placed once. Traders' reactions to price movements like keeping an order active when the price moves closer or cancelling an order when the price moves farther away are not integrated. Orders will be cancelled after their initially set lifetime exceeds, regardless of how close or far they are to the quotes at that time.

\section{Conclusion}
\label{Conclusion}
We analyzed the lifetime and volume distribution of limit orders in the limit order book. We found that short-living limit orders are the large majority of all limit orders, but they only contribute to a sparse filling of the limit order book around the midpoint price. Almost the complete filling of the limit order book around the quotes is provided by limit orders with medium lifetimes of 10\,s $\leq \tau_i <$ 10\,min. Long-living limit orders are only a small share of all limit orders but provide almost the complete volume in the depth of the limit order book.

Hence we separate the limit order book in two price regimes with different characteristics. The first regime is the liquidity cushion around the midpoint price. It follows changes of the midpoint price closely and adapts constantly. The second regime is the fragmented distant field, where orders have a much larger average lifetime and it is thus more static in time.

We focused our further research on the characteristics of the liquidity cushion differentiating between small- and large-tick stocks. We uncovered that the cushion width is dependent on \textit{e.g.} the spread, midpoint price and number of quote changes per trading day, while it is independent of the volatility of the stocks. Based on our analysis, we proposed a stochastic model for the insertion of limit and market orders into the liquidity cushion. Our model uses empirical distributions for the insertion level relative to the midpoint price and the lifetime of limit orders. With our model we are able to reproduce the average filling of the liquidity cushion as well as the price range of the stock price and the spread distribution. Simplifications of the model neglecting different lifetimes and insertion probabilities for the limit order insertion on different price niveaus fail to reproduce the average filling of the liquidity cushion, while other characteristics like the return and spread distribution as well as the price range do not seem to be affected by this.

For further research it would be interesting to analyze the driving factors of the aforementioned characteristics. Furthermore the role of the fragmented distant field as a possible security against price shocks is to be studied deeper. 

\bibliography{bibliography}

\begin{thebibliography}{38}
\expandafter\ifx\csname natexlab\endcsname\relax\def\natexlab#1{#1}\fi
\expandafter\ifx\csname bibnamefont\endcsname\relax
  \def\bibnamefont#1{#1}\fi
\expandafter\ifx\csname bibfnamefont\endcsname\relax
  \def\bibfnamefont#1{#1}\fi
\expandafter\ifx\csname citenamefont\endcsname\relax
  \def\citenamefont#1{#1}\fi
\expandafter\ifx\csname url\endcsname\relax
  \def\url#1{\texttt{#1}}\fi
\expandafter\ifx\csname urlprefix\endcsname\relax\def\urlprefix{URL }\fi
\providecommand{\bibinfo}[2]{#2}
\providecommand{\eprint}[2][]{\url{#2}}

\bibitem[{\citenamefont{Mandelbrot}(1997)}]{Mandelbrot.1997b}
\bibinfo{author}{\bibfnamefont{B.~B.} \bibnamefont{Mandelbrot}}, in
  \emph{\bibinfo{booktitle}{{Fractals and Scaling in Finance}}}, edited by
  \bibinfo{editor}{\bibfnamefont{B.~B.} \bibnamefont{Mandelbrot}}
  (\bibinfo{publisher}{{Springer New York}}, \bibinfo{year}{1997}), pp.
  \bibinfo{pages}{371--418}.

\bibitem[{\citenamefont{Mantegna and Stanley}(1997)}]{Mantegna.1997}
\bibinfo{author}{\bibfnamefont{R.~N.} \bibnamefont{Mantegna}} \bibnamefont{and}
  \bibinfo{author}{\bibfnamefont{H.~E.} \bibnamefont{Stanley}},
  \bibinfo{journal}{{Physica A: Statistical Mechanics and its Applications}}
  \textbf{\bibinfo{volume}{239}}, \bibinfo{pages}{255} (\bibinfo{year}{1997}).

\bibitem[{\citenamefont{M{\"u}nnix et~al.}(2012)\citenamefont{M{\"u}nnix,
  Shimada, Sch{\"a}fer, Leyvraz, Seligman, Guhr, and Stanley}}]{Munnix.2012}
\bibinfo{author}{\bibfnamefont{M.~C.} \bibnamefont{M{\"u}nnix}},
  \bibinfo{author}{\bibfnamefont{T.}~\bibnamefont{Shimada}},
  \bibinfo{author}{\bibfnamefont{R.}~\bibnamefont{Sch{\"a}fer}},
  \bibinfo{author}{\bibfnamefont{F.}~\bibnamefont{Leyvraz}},
  \bibinfo{author}{\bibfnamefont{T.~H.} \bibnamefont{Seligman}},
  \bibinfo{author}{\bibfnamefont{T.}~\bibnamefont{Guhr}}, \bibnamefont{and}
  \bibinfo{author}{\bibfnamefont{H.~E.} \bibnamefont{Stanley}},
  \bibinfo{journal}{{Scientific Reports}} \textbf{\bibinfo{volume}{2}},
  \bibinfo{pages}{644} (\bibinfo{year}{2012}).

\bibitem[{\citenamefont{Chung and Lee}(2016)}]{Chung.2016}
\bibinfo{author}{\bibfnamefont{K.~H.} \bibnamefont{Chung}} \bibnamefont{and}
  \bibinfo{author}{\bibfnamefont{A.~J.} \bibnamefont{Lee}},
  \bibinfo{journal}{{Asia-Pacific Journal of Financial Studies}}
  \textbf{\bibinfo{volume}{45}}, \bibinfo{pages}{7} (\bibinfo{year}{2016}).

\bibitem[{\citenamefont{Breuer}(2020)}]{Breuer.2020}
\bibinfo{author}{\bibfnamefont{A.}~\bibnamefont{Breuer}},
  \emph{\bibinfo{title}{{An Empirical Analysis of Order Dynamics in a High
  Frequency Trading Environment}}}, vol.~\bibinfo{volume}{49} of
  \emph{\bibinfo{series}{{Studienreihe der Stiftung Kreditwirtschaft an der
  Universit{\"a}t Hohenheim}}} (\bibinfo{publisher}{{Verlag Wissenschaft {\&}
  Praxis} and {Duncker {\&} Humblot GmbH}}, \bibinfo{year}{2020}),
  \bibinfo{edition}{1st} ed.

\bibitem[{\citenamefont{Hendershott and Riordan}(2013)}]{Hendershott.2013}
\bibinfo{author}{\bibfnamefont{T.}~\bibnamefont{Hendershott}} \bibnamefont{and}
  \bibinfo{author}{\bibfnamefont{R.}~\bibnamefont{Riordan}},
  \bibinfo{journal}{{Journal of Financial and Quantitative Analysis}}
  \textbf{\bibinfo{volume}{48}}, \bibinfo{pages}{1001} (\bibinfo{year}{2013}).

\bibitem[{\citenamefont{Farmer and Foley}(2009)}]{Farmer.2009}
\bibinfo{author}{\bibfnamefont{J.~D.} \bibnamefont{Farmer}} \bibnamefont{and}
  \bibinfo{author}{\bibfnamefont{D.}~\bibnamefont{Foley}},
  \bibinfo{journal}{{Nature}} \textbf{\bibinfo{volume}{460}},
  \bibinfo{pages}{685} (\bibinfo{year}{2009}).

\bibitem[{\citenamefont{Lee et~al.}(2013)\citenamefont{Lee, Eom, and
  Park}}]{Lee.2013}
\bibinfo{author}{\bibfnamefont{E.~J.} \bibnamefont{Lee}},
  \bibinfo{author}{\bibfnamefont{K.~S.} \bibnamefont{Eom}}, \bibnamefont{and}
  \bibinfo{author}{\bibfnamefont{K.~S.} \bibnamefont{Park}},
  \bibinfo{journal}{{Journal of Financial Markets}}
  \textbf{\bibinfo{volume}{16}}, \bibinfo{pages}{227} (\bibinfo{year}{2013}).

\bibitem[{\citenamefont{Kirilenko et~al.}(2017)\citenamefont{Kirilenko, Kyle,
  Samadi, and Tuzun}}]{KIRILENKO.2017}
\bibinfo{author}{\bibfnamefont{A.}~\bibnamefont{Kirilenko}},
  \bibinfo{author}{\bibfnamefont{A.~S.} \bibnamefont{Kyle}},
  \bibinfo{author}{\bibfnamefont{M.}~\bibnamefont{Samadi}}, \bibnamefont{and}
  \bibinfo{author}{\bibfnamefont{T.}~\bibnamefont{Tuzun}},
  \bibinfo{journal}{{The Journal of Finance}} \textbf{\bibinfo{volume}{72}},
  \bibinfo{pages}{967} (\bibinfo{year}{2017}).

\bibitem[{\citenamefont{Braun et~al.}(2018)\citenamefont{Braun, Fiegen, Wagner,
  Krause, and Guhr}}]{Braun.2018}
\bibinfo{author}{\bibfnamefont{T.}~\bibnamefont{Braun}},
  \bibinfo{author}{\bibfnamefont{J.~A.} \bibnamefont{Fiegen}},
  \bibinfo{author}{\bibfnamefont{D.~C.} \bibnamefont{Wagner}},
  \bibinfo{author}{\bibfnamefont{S.~M.} \bibnamefont{Krause}},
  \bibnamefont{and} \bibinfo{author}{\bibfnamefont{T.}~\bibnamefont{Guhr}},
  \bibinfo{journal}{{PLOS ONE}} \textbf{\bibinfo{volume}{13}},
  \bibinfo{pages}{e0196920} (\bibinfo{year}{2018}).

\bibitem[{\citenamefont{Patzelt and Pawelzik}(2013)}]{Patzelt.2013}
\bibinfo{author}{\bibfnamefont{F.}~\bibnamefont{Patzelt}} \bibnamefont{and}
  \bibinfo{author}{\bibfnamefont{K.}~\bibnamefont{Pawelzik}},
  \bibinfo{journal}{{Scientific Reports}} \textbf{\bibinfo{volume}{3}},
  \bibinfo{pages}{2784} (\bibinfo{year}{2013}).

\bibitem[{\citenamefont{Meudt et~al.}(2016)\citenamefont{Meudt, Schmitt,
  Sch{\"a}fer, and Guhr}}]{Meudt.2016}
\bibinfo{author}{\bibfnamefont{F.}~\bibnamefont{Meudt}},
  \bibinfo{author}{\bibfnamefont{T.~A.} \bibnamefont{Schmitt}},
  \bibinfo{author}{\bibfnamefont{R.}~\bibnamefont{Sch{\"a}fer}},
  \bibnamefont{and} \bibinfo{author}{\bibfnamefont{T.}~\bibnamefont{Guhr}},
  \bibinfo{journal}{{Physica A: Statistical Mechanics and its Applications}}
  \textbf{\bibinfo{volume}{453}}, \bibinfo{pages}{228} (\bibinfo{year}{2016}).

\bibitem[{\citenamefont{Schmitt et~al.}(2012)\citenamefont{Schmitt,
  Sch{\"a}fer, M{\"u}nnix, and Guhr}}]{Schmitt.2012}
\bibinfo{author}{\bibfnamefont{T.~A.} \bibnamefont{Schmitt}},
  \bibinfo{author}{\bibfnamefont{R.}~\bibnamefont{Sch{\"a}fer}},
  \bibinfo{author}{\bibfnamefont{M.~C.} \bibnamefont{M{\"u}nnix}},
  \bibnamefont{and} \bibinfo{author}{\bibfnamefont{T.}~\bibnamefont{Guhr}},
  \bibinfo{journal}{{EPL (Europhysics Letters)}}
  \textbf{\bibinfo{volume}{100}}, \bibinfo{pages}{38005}
  (\bibinfo{year}{2012}).

\bibitem[{\citenamefont{Krause and Bornholdt}(2013)}]{Krause.2013}
\bibinfo{author}{\bibfnamefont{S.~M.} \bibnamefont{Krause}} \bibnamefont{and}
  \bibinfo{author}{\bibfnamefont{S.}~\bibnamefont{Bornholdt}},
  \bibinfo{journal}{{Physica A: Statistical Mechanics and its Applications}}
  \textbf{\bibinfo{volume}{392}}, \bibinfo{pages}{4048} (\bibinfo{year}{2013}).

\bibitem[{\citenamefont{Plerou et~al.}(1999)\citenamefont{Plerou, Gopikrishnan,
  {Nunes Amaral}, Meyer, and Stanley}}]{Plerou.1999}
\bibinfo{author}{\bibfnamefont{V.}~\bibnamefont{Plerou}},
  \bibinfo{author}{\bibfnamefont{P.}~\bibnamefont{Gopikrishnan}},
  \bibinfo{author}{\bibfnamefont{L.~A.} \bibnamefont{{Nunes Amaral}}},
  \bibinfo{author}{\bibfnamefont{M.}~\bibnamefont{Meyer}}, \bibnamefont{and}
  \bibinfo{author}{\bibfnamefont{H.~E.} \bibnamefont{Stanley}},
  \bibinfo{journal}{{Physical review. E, Statistical physics, plasmas, fluids,
  and related interdisciplinary topics}} \textbf{\bibinfo{volume}{60}},
  \bibinfo{pages}{6519} (\bibinfo{year}{1999}).

\bibitem[{\citenamefont{Cont}(2001)}]{Cont.2001}
\bibinfo{author}{\bibfnamefont{R.}~\bibnamefont{Cont}},
  \bibinfo{journal}{{Quantitative Finance}} \textbf{\bibinfo{volume}{1}},
  \bibinfo{pages}{223} (\bibinfo{year}{2001}).

\bibitem[{\citenamefont{Gould et~al.}(2013)\citenamefont{Gould, Porter,
  Williams, McDonald, Fenn, and Howison}}]{Gould.2013}
\bibinfo{author}{\bibfnamefont{M.~D.} \bibnamefont{Gould}},
  \bibinfo{author}{\bibfnamefont{M.~A.} \bibnamefont{Porter}},
  \bibinfo{author}{\bibfnamefont{S.}~\bibnamefont{Williams}},
  \bibinfo{author}{\bibfnamefont{M.}~\bibnamefont{McDonald}},
  \bibinfo{author}{\bibfnamefont{D.~J.} \bibnamefont{Fenn}}, \bibnamefont{and}
  \bibinfo{author}{\bibfnamefont{S.~D.} \bibnamefont{Howison}},
  \bibinfo{journal}{{Quantitative Finance}} \textbf{\bibinfo{volume}{13}},
  \bibinfo{pages}{1709} (\bibinfo{year}{2013}).

\bibitem[{\citenamefont{Theissen et~al.}(2017)\citenamefont{Theissen, Krause,
  and Guhr}}]{Theissen.2017}
\bibinfo{author}{\bibfnamefont{M.}~\bibnamefont{Theissen}},
  \bibinfo{author}{\bibfnamefont{S.~M.} \bibnamefont{Krause}},
  \bibnamefont{and} \bibinfo{author}{\bibfnamefont{T.}~\bibnamefont{Guhr}},
  \bibinfo{journal}{{The European Physical Journal B}}
  \textbf{\bibinfo{volume}{90}}, \bibinfo{pages}{685} (\bibinfo{year}{2017}).

\bibitem[{\citenamefont{Bouchaud et~al.}(2002)\citenamefont{Bouchaud,
  M{\'e}zard, and Potters}}]{Bouchaud.2002}
\bibinfo{author}{\bibfnamefont{J.-P.} \bibnamefont{Bouchaud}},
  \bibinfo{author}{\bibfnamefont{M.}~\bibnamefont{M{\'e}zard}},
  \bibnamefont{and} \bibinfo{author}{\bibfnamefont{M.}~\bibnamefont{Potters}},
  \bibinfo{journal}{{Quantitative Finance}} \textbf{\bibinfo{volume}{2}},
  \bibinfo{pages}{251} (\bibinfo{year}{2002}).

\bibitem[{\citenamefont{Potters and Bouchaud}(2003)}]{Potters.2003}
\bibinfo{author}{\bibfnamefont{M.}~\bibnamefont{Potters}} \bibnamefont{and}
  \bibinfo{author}{\bibfnamefont{J.-P.} \bibnamefont{Bouchaud}},
  \bibinfo{journal}{{Physica A: Statistical Mechanics and its Applications}}
  \textbf{\bibinfo{volume}{324}}, \bibinfo{pages}{133} (\bibinfo{year}{2003}).

\bibitem[{\citenamefont{Challet and Stinchcombe}(2001)}]{Challet.2001}
\bibinfo{author}{\bibfnamefont{D.}~\bibnamefont{Challet}} \bibnamefont{and}
  \bibinfo{author}{\bibfnamefont{R.}~\bibnamefont{Stinchcombe}},
  \bibinfo{journal}{{Physica A: Statistical Mechanics and its Applications}}
  \textbf{\bibinfo{volume}{300}}, \bibinfo{pages}{285} (\bibinfo{year}{2001}).

\bibitem[{\citenamefont{Eisler et~al.}(2009)\citenamefont{Eisler, Kert{\'e}sz,
  Lillo, and Mantegna}}]{Eisler.2009}
\bibinfo{author}{\bibfnamefont{Z.}~\bibnamefont{Eisler}},
  \bibinfo{author}{\bibfnamefont{J.}~\bibnamefont{Kert{\'e}sz}},
  \bibinfo{author}{\bibfnamefont{F.}~\bibnamefont{Lillo}}, \bibnamefont{and}
  \bibinfo{author}{\bibfnamefont{R.~N.} \bibnamefont{Mantegna}},
  \bibinfo{journal}{{Quantitative Finance}} \textbf{\bibinfo{volume}{9}},
  \bibinfo{pages}{547} (\bibinfo{year}{2009}).

\bibitem[{\citenamefont{Cont et~al.}(2010)\citenamefont{Cont, Stoikov, and
  Talreja}}]{Cont.2010}
\bibinfo{author}{\bibfnamefont{R.}~\bibnamefont{Cont}},
  \bibinfo{author}{\bibfnamefont{S.}~\bibnamefont{Stoikov}}, \bibnamefont{and}
  \bibinfo{author}{\bibfnamefont{R.}~\bibnamefont{Talreja}},
  \bibinfo{journal}{{Operations Research}} \textbf{\bibinfo{volume}{58}},
  \bibinfo{pages}{549} (\bibinfo{year}{2010}).

\bibitem[{\citenamefont{Gar{\`e}che et~al.}(2013)\citenamefont{Gar{\`e}che,
  Disdier, Kockelkoren, and Bouchaud}}]{Gareche.2013}
\bibinfo{author}{\bibfnamefont{A.}~\bibnamefont{Gar{\`e}che}},
  \bibinfo{author}{\bibfnamefont{G.}~\bibnamefont{Disdier}},
  \bibinfo{author}{\bibfnamefont{J.}~\bibnamefont{Kockelkoren}},
  \bibnamefont{and} \bibinfo{author}{\bibfnamefont{J.-P.}
  \bibnamefont{Bouchaud}}, \bibinfo{journal}{{Physical review. E, Statistical,
  nonlinear, and soft matter physics}} \textbf{\bibinfo{volume}{88}},
  \bibinfo{pages}{032809} (\bibinfo{year}{2013}).

\bibitem[{\citenamefont{Huang et~al.}(2015)\citenamefont{Huang, Lehalle, and
  Rosenbaum}}]{Huang.2015}
\bibinfo{author}{\bibfnamefont{W.}~\bibnamefont{Huang}},
  \bibinfo{author}{\bibfnamefont{C.-A.} \bibnamefont{Lehalle}},
  \bibnamefont{and}
  \bibinfo{author}{\bibfnamefont{M.}~\bibnamefont{Rosenbaum}},
  \bibinfo{journal}{{Journal of the American Statistical Association}}
  \textbf{\bibinfo{volume}{110}}, \bibinfo{pages}{107} (\bibinfo{year}{2015}).

\bibitem[{tra({\natexlab{a}})}]{tradingphysics}
\emph{\bibinfo{title}{{Tradingphysics}}}, \bibinfo{note}{accessed: May 12,
  2021},
  \urlprefix\url{http://tradingphysics.com/Resources/Specifications/HistoricalItch.aspx}.

\bibitem[{\citenamefont{Huang and Polak}(2011)}]{Huang.2011}
\bibinfo{author}{\bibfnamefont{R.}~\bibnamefont{Huang}} \bibnamefont{and}
  \bibinfo{author}{\bibfnamefont{T.}~\bibnamefont{Polak}},
  \emph{\bibinfo{title}{{LOBSTER: Limit Order Book Reconstruction System}}}
  (\bibinfo{year}{2011}).

\bibitem[{tra({\natexlab{b}})}]{tradingphysics_specs}
\emph{\bibinfo{title}{{Historical TotalView-ITCH File Specifications}}},
  \bibinfo{note}{accessed: May 12, 2021},
  \urlprefix\url{http://tradingphysics.com/Resources/Specifications/HistoricalItch.aspx}.

\bibitem[{nas()}]{nasdaq_hours}
\emph{\bibinfo{title}{{Stock Market Trading Hours for Nasdaq}}},
  \bibinfo{note}{accessed: May 12, 2021},
  \urlprefix\url{https://www.nasdaq.com/stock-market-trading-hours-for-nasdaq}.

\bibitem[{\citenamefont{Todd et~al.}(2014)\citenamefont{Todd, Scherer, Beling,
  Paddrik, and Haynes}}]{Todd.2014}
\bibinfo{author}{\bibfnamefont{A.}~\bibnamefont{Todd}},
  \bibinfo{author}{\bibfnamefont{W.}~\bibnamefont{Scherer}},
  \bibinfo{author}{\bibfnamefont{P.}~\bibnamefont{Beling}},
  \bibinfo{author}{\bibfnamefont{M.}~\bibnamefont{Paddrik}}, \bibnamefont{and}
  \bibinfo{author}{\bibfnamefont{R.}~\bibnamefont{Haynes}}, in
  \emph{\bibinfo{booktitle}{{2014 IEEE International Conference on Big Data
  (Big Data 2014)}}}, edited by
  \bibinfo{editor}{\bibfnamefont{J.}~\bibnamefont{Lin}}
  (\bibinfo{publisher}{IEEE}, \bibinfo{year}{2014}), pp.
  \bibinfo{pages}{730--735}.

\bibitem[{\citenamefont{Bookstaber et~al.}(2015)\citenamefont{Bookstaber,
  Foley, and Tivnan}}]{Bookstaber.2015}
\bibinfo{author}{\bibfnamefont{R.}~\bibnamefont{Bookstaber}},
  \bibinfo{author}{\bibfnamefont{M.}~\bibnamefont{Foley}}, \bibnamefont{and}
  \bibinfo{author}{\bibfnamefont{B.}~\bibnamefont{Tivnan}},
  \emph{\bibinfo{title}{{Market Liquidity and Heterogeneity in the Investor
  Decision Cycle}}} (\bibinfo{year}{2015}).

\bibitem[{\citenamefont{Sandoval et~al.}(2016)\citenamefont{Sandoval, Nino,
  Hernandez, and Cruz}}]{Sandoval.2016}
\bibinfo{author}{\bibfnamefont{J.}~\bibnamefont{Sandoval}},
  \bibinfo{author}{\bibfnamefont{J.}~\bibnamefont{Nino}},
  \bibinfo{author}{\bibfnamefont{G.}~\bibnamefont{Hernandez}},
  \bibnamefont{and} \bibinfo{author}{\bibfnamefont{A.}~\bibnamefont{Cruz}},
  \bibinfo{journal}{{Procedia Computer Science}} \textbf{\bibinfo{volume}{80}},
  \bibinfo{pages}{752} (\bibinfo{year}{2016}).

\bibitem[{\citenamefont{Gu et~al.}(2008{\natexlab{a}})\citenamefont{Gu, Chen,
  and Zhou}}]{Gu.2008b}
\bibinfo{author}{\bibfnamefont{G.-F.} \bibnamefont{Gu}},
  \bibinfo{author}{\bibfnamefont{W.}~\bibnamefont{Chen}}, \bibnamefont{and}
  \bibinfo{author}{\bibfnamefont{W.-X.} \bibnamefont{Zhou}},
  \bibinfo{journal}{{Physica A: Statistical Mechanics and its Applications}}
  \textbf{\bibinfo{volume}{387}}, \bibinfo{pages}{5182}
  (\bibinfo{year}{2008}{\natexlab{a}}).

\bibitem[{\citenamefont{Maslov and Mills}(2001)}]{Maslov.2001}
\bibinfo{author}{\bibfnamefont{S.}~\bibnamefont{Maslov}} \bibnamefont{and}
  \bibinfo{author}{\bibfnamefont{M.}~\bibnamefont{Mills}},
  \bibinfo{journal}{{Physica A: Statistical Mechanics and its Applications}}
  \textbf{\bibinfo{volume}{299}}, \bibinfo{pages}{234} (\bibinfo{year}{2001}).

\bibitem[{\citenamefont{Mu et~al.}(2009)\citenamefont{Mu, Chen, Kert{\'e}sz,
  and Zhou}}]{Mu.2009}
\bibinfo{author}{\bibfnamefont{G.-H.} \bibnamefont{Mu}},
  \bibinfo{author}{\bibfnamefont{W.}~\bibnamefont{Chen}},
  \bibinfo{author}{\bibfnamefont{J.}~\bibnamefont{Kert{\'e}sz}},
  \bibnamefont{and} \bibinfo{author}{\bibfnamefont{W.-X.} \bibnamefont{Zhou}},
  \bibinfo{journal}{{The European Physical Journal B}}
  \textbf{\bibinfo{volume}{68}}, \bibinfo{pages}{145} (\bibinfo{year}{2009}).

\bibitem[{\citenamefont{Gu et~al.}(2008{\natexlab{b}})\citenamefont{Gu, Chen,
  and Zhou}}]{Gu.2008}
\bibinfo{author}{\bibfnamefont{G.-F.} \bibnamefont{Gu}},
  \bibinfo{author}{\bibfnamefont{W.}~\bibnamefont{Chen}}, \bibnamefont{and}
  \bibinfo{author}{\bibfnamefont{W.-X.} \bibnamefont{Zhou}},
  \bibinfo{journal}{{Physica A: Statistical Mechanics and its Applications}}
  \textbf{\bibinfo{volume}{387}}, \bibinfo{pages}{3173}
  (\bibinfo{year}{2008}{\natexlab{b}}).

\bibitem[{\citenamefont{Maslov}(2000)}]{Maslov.2000}
\bibinfo{author}{\bibfnamefont{S.}~\bibnamefont{Maslov}},
  \bibinfo{journal}{{Physica A: Statistical Mechanics and its Applications}}
  \textbf{\bibinfo{volume}{278}}, \bibinfo{pages}{571} (\bibinfo{year}{2000}).

\bibitem[{\citenamefont{T{\'o}th et~al.}(2009)\citenamefont{T{\'o}th,
  Kert{\'e}sz, and Farmer}}]{Toth.2009}
\bibinfo{author}{\bibfnamefont{B.}~\bibnamefont{T{\'o}th}},
  \bibinfo{author}{\bibfnamefont{J.}~\bibnamefont{Kert{\'e}sz}},
  \bibnamefont{and} \bibinfo{author}{\bibfnamefont{J.~D.}
  \bibnamefont{Farmer}}, \bibinfo{journal}{{The European Physical Journal B}}
  \textbf{\bibinfo{volume}{71}}, \bibinfo{pages}{499} (\bibinfo{year}{2009}).

\end{thebibliography}

\end{document}